\documentstyle[aps,prl,epsfig,twocolumn]{revtex}

\def\be{\begin{equation}}
\def\ee{\end{equation}}
\def\bea{\begin{eqnarray}}
\def\eea{\end{eqnarray}}
\def\bma{\begin{mathletters}}
\def\ema{\end{mathletters}}
\def\C{\hbox{$\mit I$\kern-.6em$\mit C$}}

\begin{document}
\draft

\title{Optimally Conclusive Discrimination of Non-orthogonal Entangled States Locally}

\author{Yi-Xin Chen and Dong Yang}

\address{Zhejiang Institute of Modern Physics and
Department of Physics, Zhejiang University, Hangzhou 310027, P.R. China} 

\date{\today}

\maketitle

\begin{abstract}
We consider one copy of a quantum system prepared with equal prior probability in one of two non-orthogonal entangled states of multipartite distributed among separated parties. We demonstrate that these two states can be optimally distinguished in the sense of conclusive discrimination by local operations and classical communications (LOCC) alone. And this proves strictly the conjecture that Virmani et.al. \cite{Virmani} confirmed numerically and analytically. Generally the optimal protocol requires local POVM operations which are explicitly constructed. The result manifests that the distinguishable information is obtained only and completely at the last operation and all prior ones give no information about that state.    
\end{abstract}


\narrowtext

%-------------------------------------------------------------
In quantum information theory, two fascinating properties are distinguished from classical information. One is entanglement and the other is non-orthogonality. Entanglement lies at the heart of many aspects of quantum information theory, such as quantum information \cite{BNS}, quantum computation \cite{Barenco}, quantum error-correction \cite{Bennett1}, and teleportation \cite{Bennett2}. Without entanglement many quantum tasks could not be carried out. In this sense, it is a quantum resource. It is a key point that it is impossible to discriminate perfectly between non-orthogonal quantum states if only one copy is provided. The well-known no-cloning theorem \cite{Wootters} demonstrates that non-orthogonal states can not be cloned exactly. Generally, orthogonal states may be distinguished perfectly only by means of global measurements since quantum information of orthogonality may be encoded in entanglement which may not be extracted by LOCC operations. Bennett et.al. \cite{Bennett} showed that there exist bases of product orthogonal pure state which can not be locally reliably distinguished despite the fact that each state in the basis contains no entanglement. Recently, Walgate et.al. \cite{Walgate} demonstrated that any two orthogonal multipartite pure states can be distinguished perfectly by only LOCC operations. Virmani et.al. \cite{Virmani} utilized their result \cite{Walgate} to show that optimal discrimination of two non-orthogonal pure states can also be achieved by LOCC in the sense of inconclusive discrimination. They also numerically and analytically confirmed that it is the case for a large set of states in conclusive discrimination. 
The problem of identifying two non-orthogonal states has been considered in \cite{IDP} and \cite{JS} by global measurements. We have discussed the problem of discriminating two non-orthogonal product states locally \cite{CY}. 
In this letter, we consider the issue of conclusive discrimination of two non-orthogonal entangled states and prove strictly the conjecture that the optimal discrimination by global measurements can be achieved by LOCC operations. \\

Suppose Alice and Bob know the precise forms of two entangled states in which one of them is shared between them. These two possible entangled states, $|\phi\rangle$ and $|\psi\rangle$ generally non-orthogonal are provided with equal prior probability. They are separated from each other and can communicate classical information only. Their aim is to identify the shared states optimally in the sense of conclusive discrimination by LOCC operations. Conclusive discrimination means that our measurement on the copy gives three outcomes which allow us to determine the prior state is $|\phi\rangle$ or $|\psi\rangle$ with certainty or "don't know". The optimization of conclusive discrimination is to obtain the maximal probability of decisive outcomes. $|\phi\rangle$ and $|\psi\rangle$ can be represented in general form: 

\bea
|\phi\rangle=\sum_{i=1}^{n}{\sqrt{r_{i}}|e_{i}\rangle_{A}|\eta_{i}\rangle_{B}},\nonumber\\
|\psi\rangle=\sum_{i=1}^{n}{\sqrt{s_{i}}|e_{i}\rangle_{A}|\gamma_{i}\rangle_{B}},
\eea
where $\{|e_{i}\rangle_{A}\}$ form an orthonormal basis set for Alice, and the vectors $\{|\eta_{i}\rangle_{B}\}$ and $\{|\gamma_{i}\rangle_{B}\}$ are normalized and generally non-orthogonal. In \cite{Walgate}, it was proved that the two states can be expressed as the following form in another orthonormal basis set on Alice's side:
\bea
|\phi\rangle=\sum_{i=1}^{n}{\sqrt{r_{i}^{'}}|e_{i}^{'}\rangle_{A}|\eta_{i}^{'}\rangle_{B}},\nonumber\\
|\psi\rangle=\sum_{i=1}^{n}{\sqrt{s_{i}^{'}}|e_{i}^{'}\rangle_{A}|\gamma_{i}^{'}\rangle_{B}},
\eea
satisfying 
\be
\sqrt{r_{i}^{'}s_{i}^{'}}\langle\eta_{i}^{'}|\gamma_{i}^{'}\rangle_{B}=\sqrt{r_{j}^{'}s_{j}^{'}}\langle\eta_{j}^{'}|\gamma_{j}^{'}\rangle_{B},
\ee
where $\{|e_{i}^{'}\rangle_{A}\}$ forms another orthonormal basis set. For orthogonal states, Walgate et.al. showed that $\langle\eta_{i}^{'}|\gamma_{i}^{'}\rangle_{B}=0$ for all $i=1,2, \cdots, n$ and proved that Alice and Bob can always distinguish between the two possible orthogonal states perfectly by LOCC operations. In the following, we suppose that the two states have been expressed as the form above and denote them still as their original form for convenience. Before our main theorem, let us introduce lemma 1. \\

Lemma 1  Let $M$ be $2 \times 2$ matrix $\left( {x \atop z}{y \atop t} \right)$ whose diagonal elements are real, $U$ unitary matrix $\left( {\cos \theta \atop \sin \theta e^{-i\omega}}{\sin \theta e^{i\omega} \atop -\cos \theta}\right)$. There exists $U$ such that the diagonal elements of $UMU^{\dagger}$ are real and of which this property is independent of $\theta $.

Proof: This lemma can be easily proved by direct computation. 
\bea
x^{'}=x \cos^{2} \theta +t \sin^{2} \theta +\sin \theta \cos \theta (y e^{-i \omega}+z e^{i \omega}), \nonumber\\
t^{'}=x \sin^{2} \theta +t \cos^{2} \theta -\sin \theta \cos \theta (y e^{-i \omega}+z e^{i \omega}). 
\eea
Set $Im(y e^{-i \omega}+z e^{i \omega})=0$ and there will always be an angle $\omega$ satisfying the equation which is explicitly independent of $\theta $. This completes the proof $\Box$.
Employing lemma 1, we can transform the two states further to the form that is expressed as theorem 1.
\\

Theorem 1 In a proper orthonormal basis set $\{|i\rangle\}$ on Alice's side, $|\phi\rangle$ and $|\psi\rangle$ can be expressed as the form:
\bea
|\phi\rangle=\sum_{i=1}^{n} \sqrt{t_{i}}|i\rangle|\mu_{i}\rangle, \nonumber\\
|\psi\rangle=\sum_{i=1}^{n} \sqrt{t_{i}}|i\rangle|\nu_{i}\rangle,
\eea  
and $|\mu_{i}\rangle$, $|\nu_{i}\rangle$ satisfy the condition that the phase difference between  each $\langle\mu_{i}|\nu_{i}\rangle$ and $\langle\phi|\psi\rangle$ is 0 or $\pi$.

Proof: Suppose $\langle\phi|\psi\rangle$ is real and we will show this does not lose any generality for the complex case. We also suppose that $|\phi\rangle$ and $|\psi\rangle$ have been expressed as the form of $(1)$ and satisfy 
$\sqrt{r_{i}s_{i}}\langle\eta_{i}|\gamma_{i}\rangle_{B}=\sqrt{r_{j}s_{j}}\langle\eta_{j}|\gamma_{j}\rangle_{B}$. It's explicit that every $\langle\eta_{i}|\gamma_{i}\rangle$ is real. As $\sum_{i}r_{i}=\sum_{i}s_{i}=1$, there must exist $r_{i}, s_{i}$ and $r_{j}, s_{j}$
satisfying $r_{i}\ge s_{i}, r_{j}\le s_{j}$. Without no loss of generality, we set $r_{1}\ge s_{1}, r_{2}\le s_{2}$. We first change the two basis $\{|e_{1}\rangle, |e_{2}\rangle\}$ into $\{|e_{1}^{'}\rangle, |e_{2}^{'}\rangle\}$ only. According the result in \cite{HJW}, the corresponding terms on Bob's side transform as:
\bea
\left( {\cos\theta \atop \sin\theta e^{i\omega}} {\sin \theta e^{-i\omega} \atop -\cos \theta} \right) \left( {\sqrt{r_{1}}|\eta_{1}\rangle \atop \sqrt{r_{2}}|\eta_{2}\rangle} \right)=\left( {\sqrt{r_{1}^{'}}|\eta_{1}^{'}\rangle \atop \sqrt{r_{2}^{'}}|\eta_{2}^{'}\rangle} \right), \nonumber\\
\left( {\cos\theta \atop \sin\theta e^{i\omega}} {\sin \theta e^{-i\omega} \atop -\cos \theta} \right) \left( {\sqrt{s_{1}}|\gamma_{1}\rangle \atop   \sqrt{s_{2}}|\gamma_{2}\rangle} \right)=\left( {\sqrt{s_{1}^{'}}|\gamma_{1}^{'}\rangle \atop   \sqrt{s_{2}^{'}}|\gamma_{2}^{'}\rangle} \right).
\eea 
And    
\bea
&r&_{1}^{'}=r_{1}\cos^{2}\theta+r_{2}\sin^{2}\theta \nonumber\\
&+&\sqrt{r_{1}r_{2}}\cos\theta\sin\theta(e^{-i\omega}\langle\eta_{1}|\eta_{2}\rangle+e^{i\omega}\langle\eta_{2}|\eta_{1}\rangle), \nonumber\\
&s&_{1}^{'}=s_{1}\cos^{2}\theta+s_{2}\sin^{2}\theta \nonumber\\
&+&\sqrt{s_{1}s_{2}}\cos\theta\sin\theta(e^{-i\omega}\langle\gamma_{1}|\gamma_{2}\rangle+e^{i\omega}\langle\gamma_{2}|\gamma_{1}\rangle).
\eea
The matrix $M=\left({\sqrt{r_{1}s_{1}}\langle\eta_{1}|\gamma_{1}\rangle \atop \sqrt{r_{2}s_{1}}\langle\eta_{2}|\gamma_{1}\rangle}{\sqrt{r_{1}s_{2}}\langle\eta_{1}|\gamma_{2}\rangle \atop \sqrt{r_{2}s_{2}}\langle\eta_{2}|\gamma_{2}\rangle}\right)$ is transformed to $U^{\ast}MU^{\dagger\ast}$ \cite{Walgate}. In lemma 1, we see the property that diagonal elements are real is dependent only on $\omega$ and independent of $\theta$. So the value of $\omega$ is determined by real diagonal elements. Explicitly, its solution is given by equation
\be
Im(\sqrt{r_{1}s_{2}}\langle\eta_{1}|\gamma_{2}\rangle e^{-i \omega}+\sqrt{r_{2}s_{1}}\langle\eta_{2}|\gamma_{1}\rangle e^{i \omega})=0.
\ee   
So $\langle\eta_{1}^{'}|\gamma_{1}^{'}\rangle$ and $\langle\eta_{2}^{'}|\gamma_{2}^{'}\rangle$ are real, positive or negative. Then we suppose $r_{1}^{'}=s_{1}^{'}=t_{1}$ and see whether this equation has always a solution. Denote $e^{-i\omega}\langle\eta_{1}|\eta_{2}\rangle+e^{i\omega}\langle\eta_{2}|\eta_{1}\rangle=x$, $e^{-i\omega}\langle\gamma_{1}|\gamma_{2}\rangle+e^{i\omega}\langle\gamma_{2}|\gamma_{1}\rangle=y$ for short which are real. The equation is reduced as
\bea
[(r_{1}-s_{1})+(r_{2}-s_{2})]+[(r_{1}-s_{1})-(r_{2}-s_{2})] \cos 2 \theta \nonumber\\
+(x\sqrt{r_{1}r_{2}}-y\sqrt{s_{1}s_{2}})\sin2\theta=0.
\eea
Denote $(r_{1}-s_{1})+(r_{2}-s_{2})=C$, $(r_{1}-s_{1})-(r_{2}-s_{2})=A$, $x\sqrt{r_{1}r_{2}}-y\sqrt{s_{1}s_{2}}=B$. We know $|A|\ge|C|$ from $r_{1}\ge s_{1}, r_{2}\le s_{2}$ and the equation has always a solution 
\be
\theta=-\frac{1}{2}(\arcsin\frac{C}{\sqrt{A^{2}+B^{2}}}+\arctan\frac{A}{B}).
\ee   
We notice the fact that $r_{1}+r_{2}=r_{1}^{'}+r_{2}^{'}$ under the unitary operation, so $r_{1}^{'}, r_{2}^{'}$ are also probabilities. So are $s_{1}^{'}, s_{2}^{'}$. Now we have find that in the new basis set $\{|e_{1}^{'}\rangle, |e_{2}^{'}\rangle, |e_{i}\rangle, i=3, \cdots, n\}$, the two states $|\phi\rangle, |\psi\rangle$ can be expressed as
\bea
|\phi\rangle=\sqrt{t_{1}}|e_{1}^{'}\rangle|\eta_{1}^{'}\rangle+\sqrt{r_{2}^{'}}|e_{2}^{'}\rangle|\eta_{2}^{'}\rangle 
+\sum_{i=3}^{n}{\sqrt{r_{i}}|e_{i}\rangle|\eta_{i}\rangle},\nonumber\\
|\psi\rangle=\sqrt{t_{1}}|e_{1}^{'}\rangle|\gamma_{1}^{'}\rangle+\sqrt{s_{2}^{'}}|e_{2}^{'}\rangle|\gamma_{2}^{'}\rangle 
+\sum_{i=3}^{n}{\sqrt{s_{i}}|e_{i}\rangle|\gamma_{i}\rangle},
\eea 
where all inner products of the corresponding terms remain real.
By repeating the above process for the $n-1$ terms, we could obtain the form expressed by theorem 1. It is clear that it is also the case when $\langle\phi|\psi\rangle$ is complex. What differs in real case is that the phase of inner product of each corresponding terms is equal to or $\pi$ different from that of $\langle\phi|\psi\rangle$. That completes our proof $\Box$. 
\\

In \cite{IDP} and \cite{JS}, it is proved that the optimal conclusive discrimination of two non-orthogonal states is given by $P=1-|\langle\phi|\psi\rangle|$ without any limitation of operations. For discriminating general states by LOCC operations, a restricted protocol is suggested in \cite{Virmani} that Alice performs local one-dimensional projections which would give her no information and leave Bob's particle in residual states which could perhaps be easily distinguished from each other. In our notation, these amount to that $r_{i}=s_{i}$ and $P^{L}=1-\sum_{i}|\langle\eta_{i}|\gamma_{i}\rangle|$ while the optimal discrimination is  $P^{opt}=1-|\sum_{i}\langle\eta_{i}|\gamma_{i}\rangle|$. If all the equations in addition to $P^{L}=P^{opt}$ are satisfed, then the protocol is optimal. Our main idea is simlar to theirs and our conclusion demonstrates the idea is very illuminating. However, two main obstacles are in the way, one is how to realize the equal probability of corresponding terms, the other is how to adjust the phases of all the inner products of corresponding terms to the same one. Each of them is not straightforward. To satisfy both the conditions at the same time, POVM on Alice's side is required in general. In the following theorem, we try to solve the problem.
\\

Theorem 2  Optimally conclusive discrimination between two non-orthogonal entangled states can be achieved by LOCC operations. 

Proof: In theorem 1, $|\phi\rangle, |\psi\rangle$ can be expressed as the form of equation (5) and satisfy the condition that the phase of each term $\langle\mu_{i}|\nu_{i}\rangle$ is the same as that of $\langle\phi|\psi\rangle$ or has $\pi$ difference from that of $\langle\phi|\psi\rangle$. 

If all the phases of $\langle\mu_{i}|\nu_{i}\rangle, i=1, \cdots, n$ are the same as that of $\langle\phi|\psi\rangle$, then Alice performs standard measurement on the basis set $\{|i\rangle\}$ and leaves Bob's state as $|\mu_{i}\rangle$ or $|\nu_{i}\rangle$ when $|i\rangle$ occurs. Bob performs the optimal conclusive discrimination between $|\mu_{i}\rangle$ and $|\nu_{i}\rangle$ which gives the optimal probability $P_{|i}=1-|\langle\mu_{i}|\nu_{i}\rangle|$. The overall optimal probability is averaged as 
\bea
P^{L}&=&\sum_{i}{t_{i}P_{|i}} \nonumber\\
&=&1-\sum_{i}{t_{i}|\langle\mu_{i}|\nu_{i}\rangle|} \nonumber\\
&=&1-|\sum_{i}{t_{i}\langle\mu_{i}|\nu_{i}\rangle}| \nonumber\\  
&=&1-|\langle\phi|\psi\rangle|.
\eea
The third equality comes from the same phase of $\langle\mu_{i}|\nu_{i}\rangle, i=1, \cdots, n$. And the optimal discrimination could be realized by LOCC operations. 

If there exist some terms of $\langle\eta_{i}|\gamma_{i}\rangle$ whose phases have $\pi$ difference from that of $\langle\phi|\psi\rangle$, then POVM or auxiliary system is necessarily introduced on Alice's side. Our idea is that after Alice's subsystem interacts properly with the auxiliary system $S$ on her side, the two states including auxiliary system $S$ can be expressed as 
\bea
U^{AS}|s_{0}\rangle|\phi\rangle=\sum_{i=1}^{m}\sqrt{t_{i}}|s_{i}\rangle|\phi_{i}\rangle 
+\sum_{i=m+1}^{N}\sqrt{t_{i}}|s_{i}\rangle|i\rangle|\mu_{i}\rangle, \nonumber\\
U^{AS}|s_{0}\rangle|\psi\rangle=\sum_{i=1}^{m}\sqrt{t_{i}}|s_{i}\rangle|\psi_{i}\rangle 
+\sum_{i=m+1}^{N}\sqrt{t_{i}}|s_{i}\rangle|i\rangle|\nu_{i}\rangle,
\eea
where $\langle\phi_{i}|\psi_{i}\rangle_{AB}=0$ and $\langle\mu_{i}|\nu_{i}\rangle_{B}$ have the same phase as that of $\langle\phi|\psi\rangle_{AB}$. Once we can express them as the form of equation (13), we could obtain the optimal protocol achieved by LOCC operations. If it is true, Alice can first project system $S$ onto the orthonormal basis $\{|s_{i}\rangle\}$. Occurrence of $|s_{i}\rangle, i\le m$ projects system $AB$ onto $|\phi_{i}\rangle$ or $|\psi_{i}\rangle$ which is orthogonal to each other and can be distinguished with certainty by the protocol in \cite{Walgate}. Occurrence of $|s_{i}\rangle, i>m$ projects onto $|i\rangle|\mu_{i}\rangle$ or $|i\rangle|\nu_{i}\rangle$ which can be identified conclusively on Bob's side with optimal probability $P_{|i}=1-|\langle\mu_{i}|\nu_{i}\rangle|$. And the optimal probability overall by LOCC is 
\bea
P^{L}&=&\sum_{i=1}^{m}t_{i}+\sum_{i=m+1}^{N}t_{i}(1-|\langle\mu_{i}|\nu_{i}\rangle|) \nonumber\\
&=&1-\sum_{i=m+1}^{N}t_{i}|\langle\mu_{i}|\nu_{i}\rangle| \nonumber\\
&=&1-|\sum_{i=m+1}^{N}t_{i}\langle\mu_{i}|\nu_{i}\rangle| \nonumber\\
&=&1-|\langle\phi|\psi\rangle|.
\eea 
    
In the following, we will prove that we can really transform to equation (13). Without loss of any generality, we suppose that $\langle\phi|\psi\rangle$ is real and $\langle\phi|\psi\rangle \ge 0$. Moreover, set $\langle\mu_{1}|\nu_{1}\rangle >0$ and $\langle\mu_{2}|\nu_{2}\rangle <0$. First, we deal with these two terms and choose $U_{1}^{AS}$ such that
\bea
U_{1}^{AS}|s_{0}\rangle|\phi\rangle&=&\sqrt{t_{1}}|\chi\rangle_{AS}|\mu_{1}\rangle+\sqrt{t_{2}}|\chi^{\perp}\rangle_{AS}|\mu_{2}\rangle \nonumber\\ 
&+&\sum_{i=3}^{n}{\sqrt{t_{i}}|s_{i}\rangle|i\rangle|\mu_{i}\rangle}, \nonumber\\
U_{1}^{AS}|s_{0}\rangle|\psi\rangle&=&\sqrt{t_{1}}|\chi\rangle_{AS}|\nu_{1}\rangle+\sqrt{t_{2}}|\chi^{\perp}\rangle_{AS}|\nu_{2}\rangle \nonumber\\
&+&\sum_{i=3}^{n}{\sqrt{t_{i}}|s_{i}\rangle|i\rangle|\nu_{i}\rangle},
\eea
where $\{|s_{i}\rangle, i=1, \cdots, n\}$ is a orthonormal basis set and $|\chi\rangle_{AS}$ and $|\chi^{\perp}\rangle_{AS}$ lie in the subspace spanned by $\{|s_{i}\rangle|j\rangle, i, j=1, 2\}$. our task is to find suitable forms of $|\chi\rangle_{AS}$ and $|\chi^{\perp}\rangle_{AS}$. This also means that we select proper interaction between system $AS$. We find that:  
if $t_{1}|\langle\mu_{1}|\nu_{1}\rangle|\ge t_{2}|\langle\mu_{2}|\nu_{2}\rangle|$, then we can choose
\bea
|\chi\rangle&=&\cos{\alpha}|s_{1}\rangle|1\rangle+\sin{\alpha}|s_{2}\rangle|2\rangle, \nonumber\\
|\chi^{\perp}\rangle&=&|s_{1}\rangle|2\rangle.
\eea  
The reason to choose such forms is that we want the state of $AB$ in the second term to be product vector. Substituting $\{|\chi\rangle_{AS}, |\chi^{\perp}\rangle_{AS}\}$ with the equation (15), we can get
\bea
&U&_{1}^{AS}|s_{0}\rangle|\phi\rangle=|s_{1}\rangle(\sqrt{t_{1}}\cos\alpha|1\rangle|\mu_{1}\rangle+\sqrt{t_{2}}|2\rangle|\mu_{2}\rangle) \nonumber\\
&+&|s_{2}\rangle\sqrt{t_{1}}\sin\alpha|2\rangle|\mu_{1}\rangle+\sum_{i=3}^{n}{\sqrt{t_{i}}|s_{i}\rangle|i\rangle|\mu_{i}\rangle}, \nonumber\\
&U&_{1}^{AS}|s_{0}\rangle|\psi\rangle=|s_{1}\rangle(\sqrt{t_{1}}\cos\alpha|1\rangle|\nu_{1}\rangle +\sqrt{t_{2}}|2\rangle|\nu_{2}\rangle) \nonumber\\
&+&|s_{2}\rangle\sqrt{t_{1}}\sin\alpha|2\rangle|\nu_{1}\rangle+\sum_{i=3}^{n}{\sqrt{t_{i}}|s_{i}\rangle|i\rangle|\nu_{i}\rangle}.
\eea 
It's clear that the corresponding terms remain the same probabilities. Our aim is to make the vectors of system $AB$ in the first corresponding terms orthogonal which gives equation
\be
t_{1}\cos^{2}\alpha\langle\mu_{1}|\nu_{1}\rangle+t_{2}\langle\mu_{2}|\nu_{2}\rangle=0.
\ee 
And from the supposition that $\langle\mu_{1}|\nu_{1}\rangle >0, \langle\mu_{2}|\nu_{2}\rangle <0$ and $t_{1}|\langle\mu_{1}|\nu_{1}\rangle|\ge t_{2}|\langle\mu_{2}|\nu_{2}\rangle|$, we can see it always has a solution\be
\alpha=\arccos\sqrt{-\frac{t_{2}\langle\mu_{2}|\nu_{2}\rangle}{t_{1}\langle\mu_{1}|\nu_{1}\rangle}}.
\ee
And inner product of the second corresponding terms of $AB$ has the same phase as that of $\langle\phi|\psi\rangle$. So we eliminate one negative term. If for all the negative terms we can find corresponding positive terms satisfying the above conditions, repeat the process for each pair terms and we can resolve all the negative terms and transform to the desired form. If for the negative term we cannot find its corresponding term satisfying the conditions, we can exchange the role of negative and positive terms. In this case, $\langle\mu_{1}|\nu_{1}\rangle <0, \langle\mu_{2}|\nu_{2}\rangle >0$ and $t_{1}|\langle\mu_{1}|\nu_{1}\rangle|\ge t_{2}|\langle\mu_{2}|\nu_{2}\rangle|$. We adopt the same protocol and the only difference is that the second term is negative. However, the absolute value of negative $t_{1}\langle\mu_{1}|\nu_{1}\rangle$ decreases to $|t_{1}\sin^{2}\alpha\langle\mu_{1}|\nu_{1}\rangle|$. And we can continue to reduce the absolute value of the negative term till it is transformed to positive. And we can always do that as $\langle\phi|\psi\rangle >0$ means that the sum of positive terms is larger than that of negative ones. So we can indeed obtain the form of equation (13) and can achieve the optimal discrimination by LOCC alone. In our discussion, it's easy to see this is also the case for complex $\langle\phi|\psi\rangle$. That completes our proof $\Box$.
\\
    
We have considered only the bipartite case so far, but our protocol can be easily generalized to two multipartite entangled states. As for the case of tripartite, we can group system $BC$ as one and apply the protocol between $A$ and $BC$ to transform as equation (13),
\bea
U^{AS}|s_{0}\rangle|\phi\rangle_{ABC}&=&\sum_{i=1}^{m}\sqrt{t_{i}}|s_{i}\rangle|\phi_{i}\rangle_{ABC} \nonumber\\ 
&+&\sum_{i=m+1}^{N}\sqrt{t_{i}}|s_{i}\rangle|i\rangle_{A}|\mu_{i}\rangle_{BC}, \nonumber\\
U^{AS}|s_{0}\rangle|\psi\rangle_{ABC}&=&\sum_{i=1}^{m}\sqrt{t_{i}}|s_{i}\rangle|\psi_{i}\rangle_{ABC} \nonumber\\ 
&+&\sum_{i=m+1}^{N}\sqrt{t_{i}}|s_{i}\rangle|i\rangle_{A}|\nu_{i}\rangle_{BC},
\eea
where $\langle\phi_{i}|\psi_{i}\rangle_{ABC}=0$ and $\langle\mu_{i}|\nu_{i}\rangle_{BC}$ have the same phase as that of $\langle\phi|\psi\rangle_{ABC}$. Each pair $|\phi_{i}\rangle_{ABC}, |\psi_{i}\rangle_{ABC}$, can be exactly distinguished \cite{Walgate}, while each pair $|\mu_{i}\rangle_{BC}, |\nu_{i}\rangle_{BC}$ can be optimally discriminated by $BC$ with $P_{|i}^{L}=1-|\langle\mu_{i}|\nu_{i}\rangle_{BC}|$. And averaging over all the possible cases gives the overall probability $P^{L}=1-|\langle\phi|\psi\rangle|$ that is optimal. It's noticeable that the optimal conclusive discrimination can be achieved by LOCC in the condition that in general, the operation performed by the last one provides the distinguishable information while all operations performed beforehand give no information about of $|\phi\rangle$ and $|\psi\rangle$. The operations in advance help the last one to distinguish states optimally.
\\

In conclusion, we have found the LOCC protocol achieving the optimal conclusive discrimination between two non-orthogonal entangled states occurring with equal prior probability. Generally, local POVM operations are required. Interestingly, the protocol shows that the distinguishable information is obtained at the last operation and all the ones beforehand give no information. The result strongly implies that optimal discriminatin is also achieved by LOCC for unequal prior probability. But in such situations the idea that the prior operations give no information does not work, and much more intricate transformation is needed which we will discuss in forthcoming paper.
\\
   
D. Yang thanks S. J. Gu and H. W. Wang for helpful discussion.
The work is supported by the NNSF of China (Grant No.19875041), the special 
NSF of Zhejiang Province (Grant No.RC98022) and Guang-Biao Cao Foundation in 
Zhejiang University.
%-------------------------------------------------------------

\end{document}